\documentclass[aps,prd,preprintnumbers,superscriptaddress,nofootinbib]{revtex4}%

\usepackage[dvipdfmx]{graphicx}
\usepackage{amsmath, amssymb}
\usepackage{comment}
\usepackage{color}
\usepackage{float}
\usepackage{braket}
\usepackage{ulem}

\makeatletter
    
    \@addtoreset{equation}{section}
\makeatother

\newcommand{\cK}{{\cal K}}

\newcommand{\cE}{{\cal E}}

\newcommand{\ba}{\begin{eqnarray}}
\newcommand{\ea}{\end{eqnarray}}

\newcommand{\tildec}{{\tilde c}}

\newcommand{\kminv}{k_{\rm min,V}}

\newcommand{\rhomg}{{\hat\rho_{m,g}}}
\newcommand{\rhomf}{{\hat\rho_{m,f}}}
\newcommand{\kxi}{\kappa \xi_c^2}

\newcommand{\simgt}{\lower.5ex\hbox{$\; \buildrel > \over \sim \;$}}
\newcommand{\simlt}{\lower.5ex\hbox{$\; \buildrel < \over \sim \;$}}

\def\({\biggl(}
\def\){\biggr)}
\def\[{\biggl[}
\def\]{\biggr]}
\def\bfp{{\bf p}}
\def\bfk{{\bf k}}
\def\bfx{{\bf x}}

\begin{document}

\preprint{YITP-16-60, KUNS-2626}

\title{Constraint on ghost-free bigravity from gravitational Cherenkov radiation

}

\author{Rampei Kimura}
\affiliation{Department of Physics, Tokyo Institute of Technology,
2-12-1 Ookayama, Meguro-ku, Tokyo 152-8551, Japan}

\author{Takahiro Tanaka}
\affiliation{Department of Physics, Kyoto University, 606-8502, Kyoto, Japan}
\affiliation{Yukawa Institute for Theoretical Physics,
	Kyoto University, 606-8502, Kyoto, Japan }
	
\author{Kazuhiro Yamamoto}
\affiliation{Department of Physical Sciences, Hiroshima University, Higashi-hiroshima, Kagamiyama 1- 3-1, 739-8526, Japan }

\author{Yasuho Yamashita}
\affiliation{Yukawa Institute for Theoretical Physics,
	Kyoto University, 606-8502, Kyoto, Japan }

\begin{abstract} 
We investigate gravitational Cherenkov radiation in a healthy branch of
 background solutions in the ghost-free bigravity model.
In this model, because of the modification of dispersion relations, each polarization mode can possess subluminal phase velocities, 
and the gravitational Cherenkov radiation could be potentially emitted from a relativistic particle. 
In the present paper, 
we derive conditions for the process of the gravitational Cherenkov radiation to occur and estimate
 the energy emission rate for each polarization mode.
We found that the gravitational Cherenkov radiation emitted even from an ultrahigh energy cosmic ray is sufficiently suppressed for the graviton's effective mass less than $100\,{\rm eV}$, and
the bigravity model with dark matter coupled to the hidden metric is therefore consistent with observations of high energy cosmic rays.

\end{abstract}

\maketitle

\section{Introduction}
The LIGO detection of gravitational wave signal from a pair of merging black holes
finally proved the propagation of gravitational waves \cite{:2016aa}, and 
it was reported that the Einstein theory of gravity is consistent with gravitational wave observations 
with high accuracy \cite{Collaboration:2016aa} in addition to solar-system tests \cite{Will:aa}.
On the other hand, theoretical and observational evidences imply that 
the universe is undergoing a phase of accelerated expansion
at the present epoch,
and one has to introduce an energy component with negative pressure, dubbed as dark energy,
to describe our universe \cite{Perlmutter:aa,Riess:aa}. 
Recently, modifications of Einstein's gravity 
have attracted considerable attention
as a substitute of dark energy and
have been
investigated in many literatures (see for reviews {\it e.g.}  \cite{Clifton:2011aa,Tsujikawa:2011aa}).

One of the simplest modifications of general relativity is to introduce 
a graviton mass to general relativity. 
This hypothetical massive graviton has been first introduced by Fierz and Pauli (FP)
in the context of linear theory, where its special structure of the mass term
prevents an additional degree of freedom from appearing 
in a flat background space time \cite{FP:1939aa}. 
One would naively expect that this linearized theory of massive gravity in the massless limit
reduces to general relativity. 
However, one gets an order-one modification of the propagator in the massless limit,
known as the van Dam-Veltman-Zakharov (vDVZ) discontinuity \cite{Z:1970aa,vDV:1970aa}
(See the recent developments in \cite{2015arXiv151206838D,2016arXiv160203721D}).
A solution of this problem 
by taking into account the nonlinear effect was proposed by Vainshtein \cite{Vainshtein:1972aa},
which is responsible for screening a scalar degree of freedom in massive graviton.
Although nonlinearities are essential to solve the van Dam-Veltman-Zakharov discontinuity,
an additional 6th degree of freedom, called Boulware-Deser ghost,
generally appears in such a theory \cite{Boulware:1972aa}.
However, it has been recently shown that the serious problem in the FP theory can be avoided by 
carefully choosing the potential, which consists of
an infinite series of interaction terms determined in such a way that it eliminates BD ghost 
at all orders in perturbation theory \cite{deRham:2010ik}. 
This infinite series of interactions can be expressed in a compact form \cite{Rham:2011aa},
referred to as the dRGT mass terms,
and the absence of BD ghost in non-perturbative description has been shown in \cite{Hassan:2011hr}.
These mass terms added to general relativity can successfully mimic the cosmological constant
in open Friedmann-Lema\^itre-Robertson-Walker (FLRW) spacetime \cite{DAmico:2011aa,Gumrukcuoglu:2011aa}, though this solution involves serious instabilities in
scalar and vector modes \cite{Gumrukcuoglu:2012ab,Gumrukcuoglu:2012aa,Felice:2012aa}.   

An extension of the dRGT theory is a bi-metric theory of gravity, 
which can be straightforwardly constructed without reintroducing BD ghost
by promoting the reference metric of the dRGT theory to a dynamical variable \cite{Hassan:2011aa,Hassan:2011ab}.
In this theory, referred to as the ghost-free bigravity, the physical degrees of freedom can be decomposed
into five from the massive spin-2 field and two from the massless spin-2 field.
Although similar type of FLRW solutions in the dRGT theory suffer from 
the catastrophic instabilities stated above,  
a new healthy branch of solutions (in the absence of matter field which couples to the second metric \cite{Felice:2013aa}
and in the presence of two matter fields each of which couples to either the first or second metric \cite{Felice:2014aa})
can be obtained  in a large fraction of the model parameter space
(See \cite{Volkov:2012,Maeda:2013aa,Strauss:2011aa,Akrami:2012aa,Akrami:2013aa,Konnig:2013aa,Comelli:2011aa}
about other cosmological solutions).
Although the bare graviton mass in this type of healthy solutions is
chosen to be larger than 
the Hubble parameter, 
one can evade the stringent constraints from Solar System tests by tuning the parameters 
in such a way that the Vainshtein radius is sufficiently large \cite{Felice:2013aa}.
Furthermore, in this background, because of the modified dispersion relations, the 
phase and group velocities for all polarization modes of graviton
deviate from the speed of light.

If a phase velocity of graviton is slower than the speed of light,
a relativistic particle emits gravitational Cherenkov radiation (GCR),
analogous to the electromagnetic Cherenkov radiation \cite{Moore:2001aa,Peters:1974,1980AnPhy.125...35C}. 
Interestingly, this GCR process can put a tight constraint 
on the phase velocity of graviton from 
the condition that the damping from GCR is not
significant for ultrahigh energy cosmic rays,
and it is confirmed to be useful in concrete examples of 
modified gravity \cite{Kimura:2011qn,Elliott:aa}, such as the new Ether-Einstein gravity \cite{Mattingly:aa,Jacobson:aa} 
and the most general second order scalar-tensor theory \cite{Horndeski:1974aa,Deffayet:2011aa,2011PThPh.126..511K}. 
For example, in the latter theory the phase velocity $c_T$ is constrained as $c-c_T < 2\times10^{-15}c$ \cite{Kimura:2011qn}, 
and most of parameter space in which at least one phase velocity is subluminal 
is not allowed because of  significant energy loss of high energy cosmic rays.
Furthermore, the authors in \cite{Kiyota:2015aa} investigated the constraints on
modified gravity theories with Lorentz-violating modified dispersion relations \cite{Mirshekari:2011aa}, 
$\omega^2=k^2 c_s^2 + m^2 + A k^\alpha $,
where $c_s$ and $m$ are the sound speed and the mass of graviton,
and $\alpha$ and $A$ are model parameters.
Although the constraint on the graviton mass is not stringent in this model,
the authors found that $\alpha$ and $A$ can be tightly constrained by observations of high energy cosmic rays.
Constraint on more general modified dispersion relations including spatial anisotropies was investigated in Ref. \cite{2015PhLB..749..551K}.
The ghost-free bigravity model could be also constrained by the same process of 
GCR, and if so, 
the model parameters should be chosen to be consistent with observations.
To this end,  in the present paper we estimate the emission rate of GCR from a relativistic particle 
and derive constraints on the ghost-free bigravity model  
from observations of high energy cosmic rays.

The rest of the present paper is organized as follows. 
In Sec.\,II we briefly review the ghost-free bigravity theory and its FLRW cosmology.
Then, in Sec.\,III we derive the emission rate of the gravitational Cherenkov radiation
of the tensor and the vector modes.
In Sec. IV we discuss consistency with  observations
of high energy cosmic rays.
Sec.\,V is devoted to conclusion.

Throughout the paper, we use units in which the speed of light and the Planck constant are unity, 
$c=\hbar =1$, and we follow the metric signature convention $(-,+,+,+)$.

\section{FLRW backgrounds}
In this section we briefly review the ghost-free bigravity model
and spatially homogeneous and isotropic cosmological solutions,
investigated in detail in \cite{Felice:2014aa}.
The action for the ghost-free bigravity is written as
\ba
S=
\frac{M_g^2}{2}\int d^4x\sqrt{-g} R[g]
+\frac{\kappa M_g^2}{2}\int d^4x\sqrt{-f} R[f]
+m^2 M_g^2\int d^4x\sqrt{-g} \sum_{i=0}^4 \alpha_i {\cal L}_i
+S_{m}[g]+S_{m}[f],
\ea
where $g_{\mu\nu}$ and $f_{\mu\nu}$ are, respectively, the
physical and the hidden metrics, 
$M_g$ is the 4-dimensional bare Planck mass for the physical metric $g_{\mu\nu}$,
$\kappa$ represents the ratio of the squared bare Planck masses for $g_{\mu\nu}$ and $f_{\mu\nu}$,
and $\alpha_i$ are dimensionless model parameters.
$S_m[g]$ $(S_m[f])$ is the action of a matter field 
that couples only to $g_{\mu\nu}$ $(f_{\mu\nu})$, 
which is referred to as $g$-matter ($f$-matter). 
The interaction Lagrangian ${\cal L}_i$ (dRGT mass terms) is given by
\begin{align}
  {\cal L}_0 &= 1 \, , \ \ 
  {\cal L}_1 = [\cK] \, , \ \ 
   {\cal L}_2 = {1\over2} \left([\cK]^2 - [\cK^2]\right) \, , \ \ 
   {\cal L}_3 = {1\over6} \left( [\cK]^3 - 3 [\cK] [\cK^2] + 2 [\cK^3] \right) \, , \\
   {\cal L}_4 &= {1\over24} \left([\cK]^4 -6 [\cK]^2 [\cK^2] + 8 [\cK] [\cK^3] +3 [\cK^2]^2 -6 [\cK^4] \right) \, ,
\end{align}
where we introduce
\ba
 {\cal K}^{\mu}_{~\nu} =\delta^\mu_{~\nu} - \bigl(\sqrt{ g^{-1} f} \bigr)^\mu_{~\nu}\,,
\ea
and $[\cK^n] = \mathrm{Tr}(\cK^n)$.
We consider the cosmological background described by the following flat FLRW metrics, 
\ba
 g_{\mu\nu} dx^\mu dx^\nu &=& -dt^2 +a^2 \delta_{ij}dx^i dx^j\,, \\
f_{\mu\nu} dx^\mu dx^\nu &=& -n^2 dt^2 +\alpha^2 \delta_{ij}dx^i dx^j,
 \label{metric}
\ea
where we set the lapse function for the physical metric to unity, 
$n=n(t)$ is the lapse function for the hidden metric, and $a=a(t)$ and $\alpha=\alpha(t)$ are 
the scale factors for the respective metrics. 
The background equations are given by 
\ba
\label{00g}
 &&3H^2 = m^2 \rhomg + {\rho_g \over M_g^2}\,, \\
 \label{00f}
 &&3H_f^2 = {m^2 \over \kappa} \rhomf + {\rho_f \over \kappa M_g^2}\,, \\
 \label{iig}
 &&2\dot{H} = m^2 \xi J(\tilde{c}-1) - {\rho_g + P_g \over M_g^2}\,, \\
 \label{iif}
 &&2{\dot{H_f} \over n} = -{m^2 \over \kappa \xi^2 \tilde{c}} \xi J(\tilde{c}-1) 
 - {\rho_f + P_f \over \kappa M_g^2}\,, 
\ea
and the energy conservation laws for $g$-matter and $f$-matter. 
Here, we defined $H\equiv \dot{a}/a$, $H_f \equiv \dot{\alpha}/(n\alpha)$, 
an overdot as the differentiation with respect to $t$, 
and $\rho_g$, $P_g$, $\rho_f$ and $P_f$ as the energy density of $g$-matter, 
the pressure of $g$-matter, the energy density of $f$-matter and the pressure of of $f$-matter, respectively. 
Also, we introduced 
\ba
 &&\rhomg \equiv U(\xi) - {\xi \over4}U'(\xi)\,,  \\
 &&\rhomf \equiv {1\over4\xi^3}U'(\xi)\,, \\
 &&  J(\xi) \equiv {1\over3} \left[ U(\xi)- {\xi \over4}U'(\xi) \right]'\,,
\ea 
where $'$ is the differentiation with respect to $\xi$ and 
\ba
 &&\xi \equiv {\alpha \over a}\,, \qquad \tilde{c} \equiv {na \over \alpha}\,, \\
 &&U(\xi) \equiv -\alpha_0+4(\xi-1)\alpha_1 -6(\xi-1)^2\alpha_2 +4(\xi-1)^3\alpha_3 - (\xi-1)^4 \alpha_4\,.
\ea
A constraint is given by the divergence of the equation of motion for $g_{\mu\nu}$  
(or equivalently by the divergence of the equation of motion for $f_{\mu\nu}$) as
\ba
 J(H-\xi H_f) =0\,.
\ea 
In this paper we focus on the healthy branch of solutions 
with $H=\xi H_f$, equivalently 
$\tildec \alpha {\dot a}-a{\dot \alpha}=0$ \cite{Felice:2013aa,Felice:2014aa}. 
Then, from Eqs.~\eqref{00g} and \eqref{00f}, we obtain 
\ba \label{energyrel}
 \rhomg(\xi)- {\xi^2 \over \kappa} \rhomf(\xi)
 =- {\rho_g \over m^2 M_g^2} + {\xi^2 \rho_f \over \kappa m^2 M_g^2}\,.
\ea
For convenience, we define $\Gamma(\xi)$ and the time-dependent effective graviton mass $\mu(\xi)$,
\ba
 \Gamma(\xi) &\equiv& \xi J(\xi) + {(\tilde{c}-1)\xi^2 \over2}J'(\xi)\,, \\
 \mu^2 (\xi)&\equiv& {1+\kappa \xi^2 \over \kappa \xi^2} m^2 \Gamma(\xi)\,. 
\ea
As is seen in the next section, $\mu$ corresponds to the effective mass 
in the long wave length limit. 

Since we are interested in the late-time cosmology,
we take the low energy limit, 
$ {\rho_g / \left(\mu^2M_g^2\right) }\ll 1$ and ${\xi^2 \rho_f / \left(\kappa \mu^2M_g^2\right) }\ll 1$. 
In this limit, from Eq.~(\ref{energyrel}), we find that  $\xi$ converges to a constant $\xi_c$ determined by
\ba
 \rhomg(\xi_c)- {\xi_c^2 \over \kappa} \rhomf(\xi_c)=0\,.
\ea
Expanding Eq.~\eqref{energyrel} around $\xi_c$, $\xi$ is given by 
\ba \label{xi-xic}
 \left( {3m^2(1+\kappa \xi_c^2)J(\xi_c) \over \kappa \xi_c} - 2\Lambda \right) {\xi -\xi_c \over \xi_c} 
 {\approx} -{\rho_g \over M_g^2} + { \xi_c^2 \rho_f \over \kappa M_g^2 },
\ea 
as a function of $\rho_g$ and $\rho_f$,
where $\Lambda$ is defined as 
\ba
 \Lambda \equiv m^2 \rhomg(\xi_c)\,.
\ea 
At least in the low energy limit, the $\xi$ parameter is monotonic, which can be seen 
from eq. (\ref{xi-xic}), since  $\rho_g$ and $\rho_f$ are also monotonic.
Then, the modified Friedmann equation for the physical metric $g_{\mu\nu}$ 
can be written as 
\ba
3H^2  \simeq {\rho_g+{\tilde \kappa}^{-1}\rho_f \over {\tilde M}_g^2} + \Lambda 
\qquad {\rm  for~}\biggl|{\Lambda \over \mu^2}\biggr| \ll 1, 
\ea
where ${\tilde M}_g^2= (1+\kappa \xi_c^2) M_g^2$
and ${\tilde \kappa} = 1 / \xi_c^4$,
and $\Lambda$ turns out to be the effective cosmological constant.

 The equation that determines the evolution of $\tildec$
 can be derived from the equation of motion for $f_{\mu\nu}$ as 
  \ba 
  {\tilde c} &=& 1+{1 \over 2W M_g^2} 
  \[\rho_g+P_g-{\tildec \xi^2 \over \kappa}(\rho_f+P_f)\] \,,
  \ea
where we define
\ba
 W &\equiv& {(1+\kappa \xi^2)J \over 2\kappa \xi} \,m^2 -H^2 \nonumber \\
 &=&{1\over2} \left( \mu^2 - {\tildec-1 \over2}
 {(1+\kappa \xi^2) J' \over \kappa} \,m^2 -2H^2 \right)\,.
\ea
In the low energy limit, we obtain 
 \ba \label{tildec}
  {\tilde c} &\simeq& 1+{1 \over M_g^2 (\mu^2-2H^2)} 
  \[\rho_g+P_g-{\xi^2 \over \kappa}(\rho_f+P_f)\] \,.
  \ea
Assuming that $W>0$, which is required to avoid the Higuchi ghost~\cite{Felice:2014aa}, 
a matter field that satisfies 
$\rho_g+P_g < \xi^2 (\rho_f+P_f)/\kappa$ implies $\tilde{c}<1$ and vice versa. 
We will show that the tensor modes of graviton possess subluminal phase velocity when $\tildec<1$ 
in Sec.~III. 
It is naively expected that cosmic-ray observations will prohibit the dominance of $f$-matter 
because it leads to the gravitational Cherenkov radiation. 
Also, we investigate the gravitational Cherenkov radiation of the vector modes of graviton. 
The vector modes can possess subluminal phase velocity for any $\tildec$, 
which will be seen in Sec.~IV. 
Therefore, even when $g$-matter dominates,  
the allowed parameter region of the ghost-free bigravity can be potentially considerably restricted.

 \section{Gravitational Cherenkov radiation of tensor modes}
 
In this section we investigate the gravitational Cherenkov radiation of the tensor modes in bigravity model.
 The tensor perturbations for the respective metrics can be introduced as small deviations from
  the background metrics (\ref{metric}), $\delta g_{ij}=a^2 (h_+ \varepsilon^+_{ij}+h_\times \varepsilon^\times_{ij})$ and $\delta f_{ij}=\alpha^2 ({\tilde h}_+ \varepsilon^+_{ij}+{\tilde h}_\times \varepsilon^\times_{ij})$, 
 where $ \varepsilon^+_{ij}$ and  $\varepsilon^\times_{ij}$
 denote the polarization tensors for plus and cross modes. 
 We normalize the polarization tensors as 
 $\varepsilon^{\mu\nu}{}^{(\lambda)}\varepsilon_{\mu\nu}^{(\lambda')}=
 \delta_{\lambda\lambda'}$.
 Hereafter we omit the index $+/\times$
 since the equations of motion are identical for both polarizations.
The quadratic action for the tensor modes is given by \cite{Felice:2014aa}
   \begin{eqnarray}
   	&&S_{\rm T} ={M_g^2\over 8}\int d^4x \biggl[{\dot h}^2-(\partial_\ell h)^2-m^2\Gamma (h-\tilde h)^2
   	+{\kappa \xi_c^2\over \tilde c}\biggl({\dot {\tilde h}}^2
   	-\tilde c^2(\partial_\ell {\tilde h})^2\biggr) \biggr].
   \end{eqnarray}
	Here, we assumed that the leading effect of non-flat background 
	is due to the deviation of $\tildec$ from unity
	and neglected the other cosmic expansion effects
\footnote{	
		One might think the other cosmic expansion effects become important at $k<H_0$. 
		However, our results will not change as long as $H_0 < \mu$.  
		Otherwise, the estimation of Eq.~(\ref{dEdtT1}) could be altered.
	}.
   Then, the equations of motion are given by
 \begin{eqnarray}
 	&&\ddot h-\triangle h +m^2 \Gamma(h-\tilde h)=0,
 	\\
 	&&\ddot {\tilde h}-{\tilde c}^2\triangle \tilde{h} +{\tildec\, m^2 \Gamma\over \kappa \xi_c^2}(\tilde h-h)=0.
 \end{eqnarray}
One can find eigen frequencies $\omega_{1,\,2}$ 
from the above equations of motion as 
\ba  \label{wavenumber1}
   { \omega_{1,\,2}^2 \over k^2} &=& 1 +{1-\tildec \over x} \left[ 1-x \mp \sqrt{(1-x)^2 + {4\kappa \xi_c^2 \over 1+\kappa \xi_c^2}x} \right]
   +\mathcal{O}\left( (1-\tildec)^2 \right)\,, 
\ea
where the upper (lower) sign is for $\omega_1$ ($\omega_2$), 
$k$ is the wave number, {\it i.e.}, $\triangle=-k^2$, and $x$ is defined as
 \footnote{  Here the sign of $x$ is different from the one in Ref.~\cite{Felice:2014aa}. }
 \begin{eqnarray}
 	x={2k^2(1-\tilde c)\over \mu^2}.
 \end{eqnarray}
 Here, the expression inside the square root in Eq.~(\ref{wavenumber1}) is always positive,
 meaning $\omega_1^2 \neq \omega_2^2$, and
 we define $\omega_{1,\, 2}$ so that the mode labeled with 1 becomes massless 
while $\omega^2_2$ reduces to $\mu^2$
 in the long wave length limit $k \rightarrow 0$.
 When $\tildec>1$, $x$ becomes negative and both modes always have 
 superluminal phase velocities, {\it i.e.}, $\omega_{1,\,2}>k$. 
 On the other hand, when $\tildec<1$, 
 the phase velocity of the mode labeled with 1 becomes subluminal 
 while that labeled with 2 is superluminal, for any $x$. 
 In order to study the gravitational Cherenkov radiation, 
 we investigate the case with $\tildec<1$, 
 in which the mode labeled with 1(2) corresponds to $\tilde{h}$ ($h$) 
 in the high energy limit $k\rightarrow \infty$ \footnote{
 In the case where $g$-matter is dominant and hence $\tildec>1$, 
 on the contrary, the the mode labeled with 1(2) reaches $h$ ($\tilde{h}$) when $k\rightarrow \infty$. }. 
 The orthogonalized action is given by 
 \begin{eqnarray}
 	&&S_{\rm T}={M_g^2\over 8}\int dt d^3k \sum_{A=1,2}\biggl(|\dot h_A|^2
 	-\omega_A^2 |h_A|^2\biggr),
 \end{eqnarray}
where the eigenfunctions $h_1$ and $h_2$ are given by
  \begin{eqnarray}
  	&&h_1=\cos\theta_g h+\sin \theta_g {\sqrt{\kappa}\xi_c \over \sqrt{\tildec}} \tilde h,
  	\\
  	&&h_2=-\sin\theta_g h+\cos \theta_g {\sqrt{\kappa}\xi_c \over \sqrt{\tildec}} \tilde h,
  \end{eqnarray}
with the mixing angle,
 \begin{eqnarray}
 	\theta_g={1\over 2}\cot^{-1}\biggl(
 	-{1+\kappa\xi_c^2\over 2\sqrt{\kappa}\xi_c\sqrt{\tildec}}x+{\tildec-\kappa \xi_c^2\over 2\sqrt{\kappa}\xi_c\sqrt{\tildec}}
 	\biggr),
 	\label{mixingAngle}
 \end{eqnarray}
 defined as a continuous function of $x$ with $0<\theta_g<\pi/2$. 

Now, we are ready to quantize the tensor modes, and the field operators can be expanded as
 \begin{eqnarray}
 	&&h_A{}_{\mu\nu}=\sqrt{4\over M_g^2}\sum_\lambda
 	\int{d^3k\over (2\pi)^{3/2}}
 	\biggl[
 	\varepsilon_{\mu\nu}^{(\lambda)}\hat a^{(\lambda)}_{A\bfk} u_{Ak}(t) e^{i{\bf k}\cdot{\bf x}}
 	+\varepsilon_{\mu\nu}^{(\lambda)}{\hat a}^{(\lambda)}_{A\bfk}{}^\dagger u_{Ak}^{*}(t) 
 	e^{-i\bfk\cdot\bfx}\biggr],
 \end{eqnarray}
 where $A=1,2$, and 
 $\hat a^{(\lambda)}_{A\bfk}{}^\dagger$ and $\hat a^{(\lambda)}_{A\bfk}$ are the creation and 
 annihilation operators, which satisfy the commutation relation 
 $[\hat a^{(\lambda)}_{A\bfk},\hat a^{(\lambda')}_{A'\bfk'}{}^\dagger]=\delta_{AA'}\delta_{\lambda\lambda'}\delta(\bfk-\bfk')
 $, 
 and the mode function 
 \begin{eqnarray}
 	&&u_{Ak} (t)={e^{-i\omega_A(k)t}\over \sqrt{2\omega_A(k)}}\,,
 \end{eqnarray}
 satisfies
 \begin{eqnarray}
 	\left({d^2\over dt^2}+\omega_A^2(k)\right)u_{Ak} (t)=0, 
 \end{eqnarray}
 and ${\dot u}^*_{Ak}(t) u_{Ak}(t)-\dot u_{Ak}(t) u_{Ak}^*(t)=i$.

 \def\bfp{{\bf p}}
 \def\bfk{{\bf k}}
 \def\bfx{{\bf x}}
 \def\hath{{\hat h}}
 \def\in{{\rm in}}
 \def\f{{\rm f}}

We are interested in the GCR from a high energy particle, {\it e.g.}, a high energy proton. 
For simplicity, we consider a complex scalar
 field with the action 
 \begin{eqnarray}
 	&&S_m=\int d^4x \sqrt{-g}\left[ - 
 	g^{\mu\nu}\partial_\mu \psi^* \partial_\nu\psi -M^2\psi^*\psi
 	\right],
 	\label{SM}
 \end{eqnarray}
 instead of a Dirac fermion. 
 Neglecting the cosmic expansion and the coupling to the metric perturbation, 
 the free part of $\psi$ can be quantized as 
 \begin{eqnarray}
 	&&\hat \psi(t,\bfx)=\int{d^3p\over (2\pi)^{3/2}}
 	\left[
 	\hat b_\bfp \psi_p(t) e^{i\bfp\cdot\bfx}
 	+\hat c_\bfp^\dagger \psi_p^*(t) e^{-i\bfp\cdot\bfx}\right],
 	\nonumber
 	\\
 \end{eqnarray}
 where $\hat b_\bfp$ and $\hat c_\bfp^\dagger$ are the annihilation and creation 
 operators of the particle and anti-particle, respectively, which satisfy 
 the commutation relations $[\hat b_\bfp,\hat b_{\bfp'}^\dagger]
 =\delta(\bfp-\bfp')$, $[\hat c_\bfp,\hat c_{\bfp'}^\dagger]=\delta(\bfp-\bfp')$,  
 and the mode function
 \begin{eqnarray}
 	\psi_p(t)={1\over \sqrt{2\Omega_p}}e^{-i\Omega_pt},
 \end{eqnarray}
  obeys
 \begin{eqnarray}
 	\left({d^2\over d t^2}+p^2+M^2 \right)\psi_p(t)=0,
 \end{eqnarray}
 with $\Omega_p=\sqrt{p^2+M^2}$. 
 The interaction part of the action (\ref{SM}) is given by
 \begin{eqnarray}
 	S_I&=&-\int dt \, d^3\!x\, h^{ij}\partial_i\psi\partial_j\psi^*,
 \end{eqnarray}
 and the interaction Hamiltonian is 
 \begin{eqnarray}
 	H_I&=& \int d^3\!x\,  h^{ij}\partial_i\psi\partial_j\psi^*.
 \end{eqnarray}
(Strictly speaking, all time derivatives must be replaced by means of
 the conjugate momenta in the Hamiltonian.)

 \begin{figure}[t]
 	\begin{center}
 		\includegraphics[width=70mm]{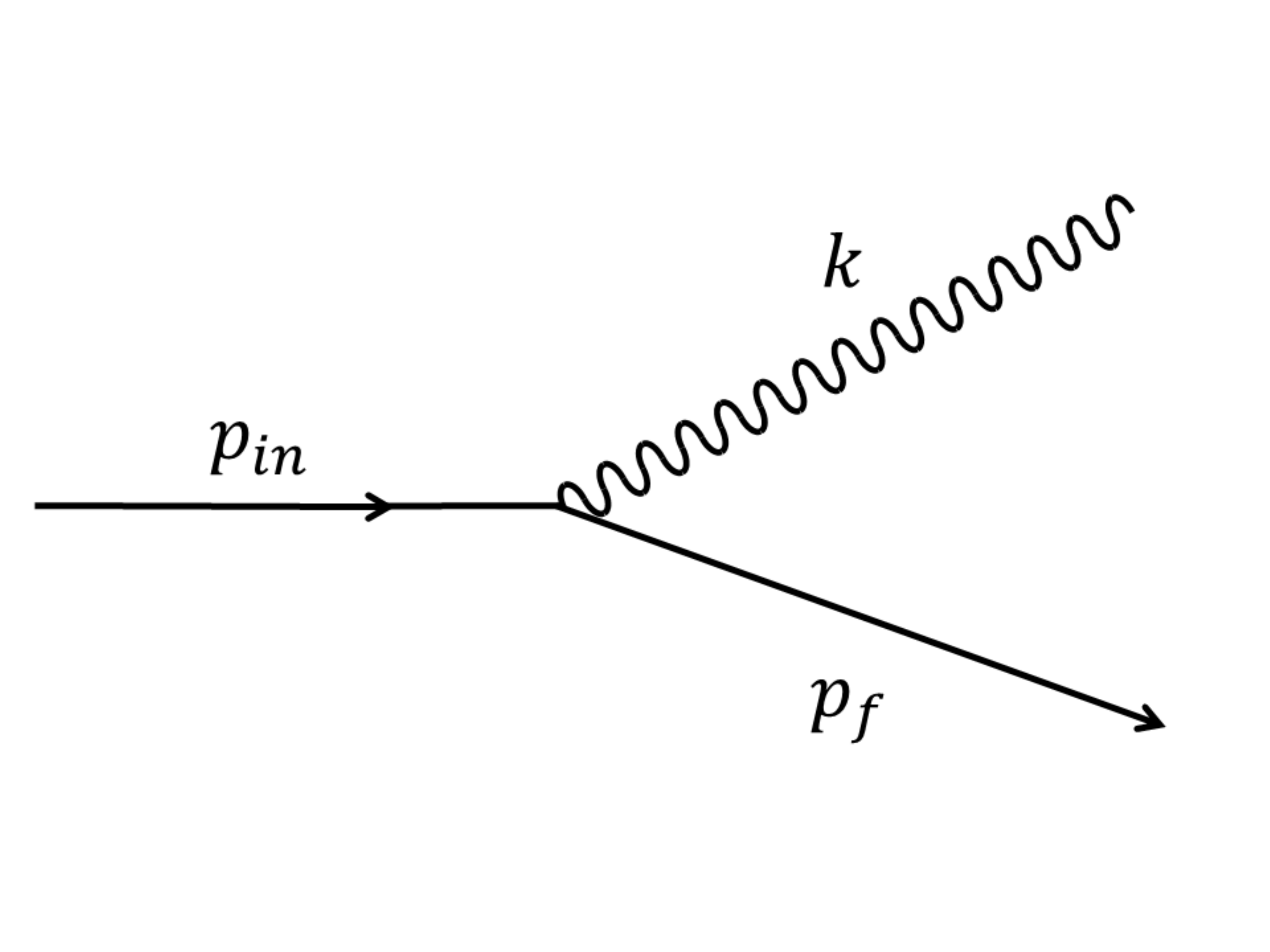}
 	\end{center}
 	\caption{Feynman diagram for the process}
 	\label{fig:one}
 \end{figure}
 
 In order to evaluate the total energy of the gravitational Cherenkov radiation, we adopt the
 method developed in \cite{Kimura:2011aa,Yamamoto:2010aa}.
 (Note that the gravitational Cherenkov radiation can be also derived classically as in the case of the electromagnetic Cherenkov radiation \cite{1980AnPhy.125...35C}.)
 Based on the in-in formalism \cite{Weinberg:aa}, at the lowest order 
 of the expectation value of the number operator of graviton is
 given by
 \begin{eqnarray}
 	&&\left<{\hat a}^{\dagger(\lambda)}_{A\bfk}{\hat a}^{(\lambda)}_{A\bfk} \right>=
 	{i^2} 
 	\int^t_{t_\in}dt_2
 	\int^{t_2}_{t_\in}dt_1
	\braket{\in |\,
	[H_I(t_1),[H_I(t_2),{\hat a}^{\dagger(\lambda)}_{A\bfk}{\hat a}^{(\lambda)}_{A\bfk} ]]
	\,| \in}
	\label{expv}
 \end{eqnarray}
for the initial state with one scalar particle with the momentum, $\bfp_\in$, {\it i.e.}, 
 $\ket{\in}={\hat b}^\dagger_{\bfp_\in}|0\rangle$. This gives the transition probability of the process
in which one graviton with the momentum $\bfk$ is emitted from a scalar particle with
 the initial momentum $\bfp_\in$ as shown in Fig.~\ref{fig:one}.
Eq.~(\ref{expv}) can be rewritten as \cite{Adshead:2009aa}
 \begin{eqnarray}
 	&&\left<{\hat a}^{\dagger(\lambda)}_{A\bfk}{\hat a}^{(\lambda)}_{A\bfk} \right>
 	=
 	2\Re\int^t_{t_\in}dt_2\int^{t_2}_{t_\in}dt_1
 	\braket{ \in | H_I(t_1)
 	{\hat a}^{\dagger(\lambda)}_{A\bfk}{\hat a}^{(\lambda)}_{A\bfk} 
 	H_I(t_2) |\in}.
 	\nonumber\\
 \end{eqnarray}
 Hereafter, we omit the tensor mode labeled with 2, whose phase velocity is always superluminal.
 Then, the total radiation energy emitted from the scalar particle
 into the tensor mode labeled with 1 can be estimated as 
 $E=\sum_\lambda\sum_\bfk \omega_k$ $\bigl<{\hat a}^{\dagger(\lambda)}_{A\bfk}
 {\hat a}^{(\lambda)}_{A\bfk} \bigr>$,
 which leads to
 \begin{eqnarray}
 	&&E_{\rm T}=
 	\int{d^3k\over (2\pi)^3}{\omega_{1}}
 	\biggl|\int_{t_\in}^t dt_1
 	\sqrt{{4\over M_g^2}}
 	u_{1k}(t_1)\psi_{p_{\rm f}}(t_1)
 	\psi_{p_\in}^*(t_1)\varepsilon_{ij}^{(\lambda)}p_{\in}^i p^j_{\rm f}
 	\biggr|^2\cos^2\!\theta_g\,,
 \end{eqnarray}
 where $\bfp_\f+\bfk=\bfp_\in~ (p_\f^i+k^i=p_\in^i)$. With the aid of the relation 
 $\sum_\lambda \bigl|\varepsilon_{ij}^{(\lambda)}p_\in^i p_\f^j\bigr|^2=p_\in^4\sin^4\!\theta/2$,
 we have
 \begin{eqnarray}
 	E_{\rm T}&=&    {1\over 2}
 	\int{d^3k\over (2\pi)^3}{\omega_{1}}p_\in^4\sin^4\!\theta
 	\biggl|
 	\int_{t_\in}^t dt_1
 	\sqrt{{4\over M_g^2}}
 	u_{1k}(t_1)
 	\psi_{p_\f}(t_1)
 	\psi_{p_\in}^*(t_1)
 	\biggr|^2\cos^2\!\theta_g\,.
 	\label{EA}
 \end{eqnarray}
After plugging the mode functions into (\ref{EA}), the total radiation energy (\ref{EA}) reduces to
 \begin{eqnarray}
 	E_{\rm T}&\simeq&{1    \over 4M_g^2}
 	\int{d^3k\over (2\pi)^3}{p_\in^4\sin^4\!\theta\over \Omega_\f\Omega_\in}
 	{2\pi T} \delta(\Omega_\in-\Omega_\f-\omega_{1})\cos^2\!\theta_g,
 \end{eqnarray}
where
$\Omega_\in=\sqrt{\bfp^2_\in+M^2}$ and 
$\Omega_\f=\sqrt{(\bfp_\in-\bfk)^2+M^2}$,
and we used 
 \begin{eqnarray}\label{approx2}
 	&&\biggl|
 	\int_{t_\in}^t dt_1
 	\exp\left[i(\Omega_\in-\Omega_\f-\omega_{1})(t_1-t_{\in})\right]
 	\biggr|^2
 	\simeq {2\pi T} \delta(\Omega_\in-\Omega_\f-\omega_{1}),
 \end{eqnarray}
 assuming the long time duration of the integration, where $T=t-t_\in$. 
 Then, we have the expression in the relativistic limit of the scalar particle, 
 $p_\in\gg M$,
 \begin{eqnarray}
 	{dE_T\over dt}&=&{p_\in^3\over 4M_g^2}\int_0^\infty{dkk^2\over 2\pi}
 	\int_{-1}^1d(\cos\theta){\sin^4\!\theta \over \Omega_\f}
 	\cos^2\!\theta_g\delta(\Omega_\in-\Omega_\f-\omega_{1}).
 	\label{dEdta}
 \end{eqnarray}
 Now, we consider the delta-function, which can be written as
 \begin{eqnarray} \label{deltaid}
 	\delta(\Omega_\in-\Omega_\f-\omega_{1})=2\Omega_\f\delta(\Omega_\f^2-(\Omega_\in-\omega_{1})^2) \Theta (\Omega_\in-\omega_{1}),
 \end{eqnarray}
 where
 $\Theta(y)$ is the Heaviside function.
 Using the identity \eqref{deltaid},
 we may write
 \begin{eqnarray}
 	&&\delta(\Omega_\in-\Omega_\f-\omega_{1})
 	={\Omega_\f\over p_\in k}\delta\biggl(\cos\theta-
	\frac{k}{2p_\in} \left(1-\frac{\omega_{1}^{2}}{k^{2}}\right) 	
 	- \sqrt{ 1+ { M^2\over \mathstrut p_\in^{2} } } \frac{\omega_{1}}{k} \biggr)
	\Theta (\Omega_\in-\omega_{1}).
 \end{eqnarray}
 Integration over $\theta$ in Eq.~(\ref{dEdta}) makes a nontrivial contribution when
 \begin{eqnarray}
 	&& 
 	\cos\theta=\frac{k}{2p_\in} \left(1-\frac{\omega_{1}^{2}}{k^{2}}\right)
 	+ \sqrt{ 1+ { M^2\over \mathstrut p_\in^{2} } } \frac{\omega_{1}}{k}
 	\leq 1,
 	\label{cosc}
 \end{eqnarray}
 which is a necessary condition for GCR to arise.
 Assuming $M/p_\in \ll 1$ and $1-\omega_1^2 /k^2 \ll1$, 
 the above condition can be rewritten as 
 \begin{eqnarray}
 	\cos\theta \approx
 	1+{M^2\over 2p_\in^2} 
	- { p_\in-k \over 2p_\in}\left(1-{\omega_1^2\over k^2}\right) \leq 1.
 	\label{cosa}
 \end{eqnarray}
 From the presence of $\Theta (\Omega_\in-\omega_{{1}})$, 
 the possible range of $k$ is restricted to $k \simeq \omega_1 \lesssim p_\in$, 
 and hence we reconfirm that $\omega_1^2<k^2$ is a necessary condition for GCR. 
 Then, the condition (\ref{cosa}) leads to
 \begin{align}
 	1-{\omega_1^2\over k^2}\geq {M^2\over p_\in \left( p_\in -k \right)}.
 	\label{GCRconditionmg}
 \end{align}
As we will see soon, the left hand side of the above inequality is 
approximated as $1-\omega_1^2 / k^2 \sim {\cal O}(1-\tildec)$ for any $k$.
Thus, the condition for emitting GCR is simply given by 
$1-\tildec \simgt M^2/p_\in^2 \sim 10^{-22}$ for a ultrahigh energy cosmic ray proton
with $p_\in \sim10^{11}{\rm GeV}$ and $M\sim1{\rm GeV}$. 
Although $M^2/p_\in^2$ term could be important when $1-\tildec \sim M^2/p_\in^2$,
the effect of mass merely reduces the GCR efficientcy.
We will find later that the constraint is always weak even 
if we neglect the proton mass to discuss the constraints from the tensor
GCR.
Therefore we can safely ignore $M^2/p_\in^2$ term in this context.
Then, the condition \eqref{GCRconditionmg} is understood 
 as the one that the effective refractive index exceeds unity,
 $n_A=k/\omega_{{A}}>1$.
 Thus, GCR is emitted only through the mode labeled with 1 when $1-\tildec>0$.
 
 We integrate Eq.~(\ref{dEdta}) by adopting the small angle approximation
 $\theta\ll1$, and we have\footnote{Contrary to the vector case, which will be seen in the next section, 
 		the condition (\ref{cosc}) does not impose the lower limit of the integration.
 		This is because the subluminal mode labeled with 1
 corresponds to the massless mode in 
the limit $k \to 0$.  }
 
 \begin{align}
 	\frac{dE_{\rm T}}{dt}={1 \over 8 \pi M_g^2}\int_{0}^{p_\in}dkk
 	\left(\left (p_\in -k \right)\left(1-{\omega_1^2\over k^2}\right)\right)^{2}\cos^2\!\theta_g.
 	\label{dEdtmg}
 \end{align}
Because of the complex $k$-dependence in $1-\omega_1^2/k^2$ and $\cos \theta_g$, 
one cannot simply integrate Eq.~(\ref{dEdtmg}).
To approximately estimate $dE_T/dt$, we consider the limiting cases with $|x| \ll 1$ and $|x| \gg 1$.
In both limits, 
$1-\omega_1^2/k^2$ is estimated from Eq.~\eqref{wavenumber1} as 
\ba
 1-{\omega_1^2 \over k^2} \simeq  \left\{ 
\begin{array}{cr}
	\displaystyle{
		 {2\kxi \over 1+\kxi} (1-\tildec) + {\cal O}(x)\,,
	} 
	&{(x \ll 1)}, \\
	\displaystyle{}\\
	\displaystyle{
		2(1-\tildec) + {\cal O}(x^{-1})\,,
	} 
	&{(x \gg 1)}.\\
	\label{condition1a}
\end{array} \right.
\ea
Also, we estimate $\cos^2\! \theta_g$ from Eq.~\eqref{mixingAngle} as 
\ba
 \cos^2\! \theta_g \simeq  \left\{ 
\begin{array}{cr}
	\displaystyle{
		 {1 \over 1+\kxi} + {\cal O}(x)\,,
	} 
	&{(x \ll 1)}, \\
	\displaystyle{}\\
	\displaystyle{
		 {\kxi \over (1+\kxi)^2} x^{-2} + {\cal O}(x^{-3})\,,
	} 
	&{(x \gg 1)}.\\
	\label{cos2thetag}
\end{array} \right.
\ea
Then, we can now estimate $dE_T/dt$ using the approximate expressions (\ref{condition1a}) and (\ref{cos2thetag}). 
Denoting the wave number at $x=1$ as $k_D \equiv \mu_c/\sqrt{2(1-\tildec)}$, 
we discuss two cases: $k_D < p_\in$ and $k_D > p_\in$, one by one.
For the case with $k_D< p_\in$, 
we can estimate $dE_T/dt$ by dividing the interval of the integral into two at $x=1$,
and we get\footnote{
		Since the dominant contribution to the integral (\ref{dEdtT1}) lies at $x \sim 1$, 
		the expression~\eqref{dEdtT1} does not smoothly connect with
		\eqref{dEdtT2} at $k_D \sim p_\in$.  
		However, it is sufficient to understand the dependence of $dE_{\rm T}/dt$ on $\mu$ and $1-\tildec$ for our present purpose.
	}
\ba	
	\frac{dE_{\rm T}}{dt} &\simeq &
	{1 \over 8 \pi M_g^2}\int_{0}^{k_D}dk
	\[ k\left(\left (p_\in-k \right)
	\left(1-{\omega_1^2\over k^2}\right)\right)^{2}\cos^2\!\theta_g\]_{|x| \ll 1}
	\nonumber\\&&
	+{1 \over 8 \pi M_g^2}\int_{k_D}^{p_\in}dk
	\[ k\left(\left (p_\in-k\right)
	\left(1-{\omega_1^2\over k^2}\right)\right)^{2}\cos^2\!\theta_g\]_{|x| \gg 1}\nonumber\\
	& \simeq &{1 \over 8 \pi M_g^2}
	{\kappa \xi_c^2 (1+2\kappa \xi_c^2) \over (1+ \kxi)^3} \,
	p_\in^2 \mu^2 (1-\tildec).
	\label{dEdtT1}
\ea
If $k_D \geq p_\in$, we only need to consider $x \ll 1 $ region, and then  $dE_T/dt$ can be approximated as
\ba
\frac{dE_{\rm T}}{dt} &\simeq&  
	{1 \over 8 \pi M_g^2}\int_{0}^{p_\in}dk
	\[ k\left(\left (p_\in-k \right)
	\left(1-{\omega_1^2\over k^2}\right)\right)^{2}\cos^2\!\theta_g\]_{|x| \ll 1}\nonumber\\
	& \simeq &
	{1 \over 8 \pi M_g^2}
{\kappa^2\xi^4_c \over 3 (1+ \kxi)^3}p_\in^4 (1-\tildec)^2.
	\label{dEdtT2}
\ea

\section{Gravitational Cherenkov radiation from vector modes}

In this section we investigate the gravitational Cherenkov radiation 
of the vector modes of graviton in the ghost-free bigravity.
We introduce vector perturbations around the cosmological background as
$\delta g_{0i}=aB_i$, $\delta f_{0i}=n\alpha b_i$, 
$\delta g_{ij}=a^2 \partial_{(i} E_{j)} $, and $\delta f_{ij}=\alpha^2 \partial_{(i} S_{j)}$,
where $B_i$, $b_i$, $E_i$ and $S_i$ are transverse vectors. 
Following the discussion in Ref.~\cite{Felice:2014aa}, 
the effective action for the vector modes is written in terms of one
dynamical vector variable for each polarization, 
while the other vectors are constrained or left unspecified corresponding to gauge degrees of freedom.   
The quadratic action for the vector mode expanded in vector harmonics is given as 
\ba
S_{\rm V} = {M_-^2\over8} \int dt d^3k a^3 A \left[ \dot{\cE}^i \dot{\cE}_i^* 
- \left\{ {k^2 \over a^2}c_{\rm V}^2 
+ m_{\rm V}^2 \right\}\cE^i \cE_i^* \right]\,,
\ea
where 
\ba
\cE_i &\equiv& \sqrt{{1+\kappa \xi^2 \over \kappa \xi^2} } k\left( E_i -S_i \right) \,, \\
\label{A}
A&\equiv& {\kappa \xi^2 \over 1+\kappa \xi^2}
\left[ {(\tilde{c}+1) {\Gamma} k^2\over 2a^2 \mu^2 {\xi J} } + {\tilde{c} +\kappa \xi^2 \over 1+\kappa \xi^2}\right]^{-1}\,, \\
M_-^2&\equiv& {\kappa \xi^2 \over 1+\kappa \xi^2} M_g^2\,, \\
c_{\rm V}^2 &\equiv& {(\tilde{c}+1) \Gamma \over 2\xi J}\,,\\
m_{\rm V}^2 &\equiv& {\tilde{c}+\kappa\xi^2 \over 1+\kappa\xi^2} \mu^2 .
\ea
$B_i$ and $b_i$ are non-dynamical degrees of freedom and written in terms of 
$E_i$ and $S_i$ by means of the constraints as 
\ba
B_i &\equiv& a \left[ {\dot{E}_i \over2} -{A\over2} \left( \dot{E}_i -\dot{S}_i \right) \right]\,, \\
b_i &\equiv& a \left[ {\dot{S}_i \over2\tilde{c}} -{A\over 2\kappa \xi_c^2} \left( \dot{E}_i -\dot{S}_i \right) \right]\,.
\ea
In the low-energy limit where $\xi \simeq \xi_c$, $c_{\rm V}^2$ is written as 
\ba
1-c_{\rm V}^2 \simeq {1-\tilde{c} \over 2} \left( 1+ {\cal C} \right)\,,
\ea
    where we define ${\cal C} \equiv \xi_c J'(\xi) / J(\xi)$.
Then, subluminal phase velocity can be achieved when $1+{\cal C}>0$ for $f$-matter dominant case $(1-\tildec>0)$ 
or  $1+{\cal C}<0$ for $g$-matter dominant case $(1-\tildec<0)$.	
Imposing the positivity of the effective mass squared,  $\mu^2>0$ and $J>0$,
and the absence of gradient instability $c_{\rm V}^2 \geq  0$,
the conditions that $c_{\rm V}$ is subluminal are given by 
\ba
\({\cal C} < -1 \cap 1< \tildec < {-2+{\cal C} \over {\cal C}} \)
\cup  
\({\cal C} > -1 \cap  {-1+{\cal C} \over 1+{\cal C}} < \tildec < 1\).
\label{conditionC}
\ea
According to \cite{Felice:2013aa}, the Vainshtein radius 
is given by 
\ba
r_{\rm V}={\cal O}\(\({|{\cal C}| r_g \over \mu^2}\)^{1/3}\,\),
\label{VainshteinRadius}
\ea
where $r_g$ is the gravitational radius of the star.
Therefore, $|{\cal C}|$ need to be sufficiently large for the Vainshtein 
mechanism to work. 
One can find such parameter spaces in the region~(\ref{conditionC})
for small $|1-\tildec|$, and 
the smallness of $|1-\tildec|$ is also consistent with 
the background equation (\ref{tildec}).
Therefore, $c_{\rm V}^2$ can be significantly subluminal 
both in the $g$-matter dominant and $f$-matter dominant cases.

Then,  we obtain the quantized vector gravitational perturbation:
\ba
\hat{\cE}_i = {1\over a} \sum_\lambda \int {d^3k \over (2\pi)^{3/2}} \sqrt{2\over A M_-^2} 
\left[ \varepsilon_i ^{(\lambda)} \hat{a}^{(\lambda)}_{{\rm V}\bfk} u_{{\rm V}k}(\eta) e^{i \bf k \cdot x}
+ \varepsilon_i ^{(\lambda)} \hat{a}^{(\lambda)}_{{\rm V}\bf k}{}^\dagger u^*_{{\rm V}k}(\eta) e^{-i \bf k \cdot x} \right]\,,
\ea
where $\varepsilon_i^{(\lambda)}$ is the polarization vector, 
which is normalized as 
$\varepsilon^{\mu}{}^{(\lambda)}\varepsilon_{\mu}^{(\lambda')}=
\delta_{\lambda\lambda'}$,
$\hat a^{(\lambda)}_{{\rm V}\bfk}{}^\dagger$ and $\hat a^{(\lambda)}_{{\rm V}\bfk}$ are the creation and 
annihilation operators, which satisfy the commutation relation 
$[\hat a^{(\lambda)}_{{\rm V}\bfk},\hat a^{(\lambda')}_{{\rm V}\bfk'}{}^\dagger]
=\delta_{\lambda\lambda'}\delta(\bfk-\bfk')$.  
Neglecting the effect of cosmic expansion and considering $a\simeq 1$, 
the mode function
\ba
 u_{{\rm V}k} (t)={e^{-i\omega_{\rm V}(k)t}\over \sqrt{2\omega_{\rm V}(k)}}\,,
\ea
satisfies
\begin{eqnarray}
	\left({d^2\over dt^2}+\omega_{\rm V}^2(k)\right)u_{{\rm V}k} (t)=0, 
\end{eqnarray}
and ${\dot u}^*_{{\rm V}k}(t) u_{{\rm V}k}(t)-\dot u_{{\rm V}k}(t) u_{{\rm V}k}^*(t)=i$ with 
\begin{eqnarray}
	\omega_{\rm V}^2(k)=c_{\rm V}^2k^2+m_{\rm V}^2. 
\end{eqnarray}

The coupling between the vector graviton and the complex scalar field $\psi$ is given as

\ba
I_{\rm int} = -\int dt\, d^3\!x\, h^{\mu\nu} \left[ \partial_\mu \psi\, \partial_\nu \psi^*
-{1\over2}  \eta_{\mu\nu} \left( \partial^\lambda \psi \partial_\lambda \psi^* +2M^2  \psi  \psi^* \right) \right]\,.
\ea
Since the whole action is invariant under a coordinate
transformation,
we impose a convenient gauge fixing condition 
\ba
S_i = {A-1 \over A} E_i\,,
\ea
so that $h_{0i}$ components vanish. 
In this gauge, $E_i$ is written in terms of $\cE_i$ as
\ba
E_i = \sqrt{ {\kappa \xi_c^2 \over 1+\kappa \xi_c^2} } {A\over k} \cE_i\,,
\ea
and the Hamiltonian for the interaction between the graviton and scalar field becomes
\ba
H_{\rm int} &=& \int d^3\!x\, {\partial}_{(i} E_{j)} \partial^i\psi \partial^j \psi^* \,.
\ea
As in Sec.\,III, we calculate 
the gravitational radiation energy emitted from the process shown in Fig.~\ref{fig:one} 
as
\ba
\label{E}
E_{\rm V}&=& \sum_\lambda \int {d^3k \over \left( 2\pi \right)^3}\, \omega_{\rm V}
\left| \int^t_{t_{ \in}} dt_1  {\sqrt{2A}\over M_g} 
u_{{\rm V}k}(t_1) \psi_{p_{\rm f}}(t_1) \psi^*_{p_{\in}}(t_1) 
{\hat{k}_{(i}} \varepsilon_{j)}^{(\lambda)} p_{\in}^i p_{\rm f}^j  \right|^2 \nonumber \\
&=&  \int {d^3k \over \left( 2\pi \right)^3} \,\omega_{\rm V}
{A\over 4M_g^2} 
p_{\in}^2 \sin^2\! \theta  \left( 2p_{\in}\cos \theta -k \right)^2
\left| \int^t_{t_{\in}} dt_1 
u_{{\rm V}k}(t_1) \psi_{p_{\rm f}}(t_1) \psi^*_{p_{\in}}(t_1) \right|^2 \,, 
\ea
where we define a unit vector parallel to $\bf k$, ${\hat{k}_i}\equiv k_i/ |{\bf k}|$, and  
$\theta$ as the angle between $\bf p_{\in}$ and $\bf k$, 
and use ${\bf p_{\in}} = {\bf k} + \bf p_{\rm f}$. 
Using Eq.~\eqref{approx1} and 
assuming the long time duration of the time integration \eqref{approx2}, 
$E_{\rm V}$ is estimated as 
\ba \label{EV}
E_{\rm V} ={t-t_{\in} \over 32\pi^2M_g^2}\int d^3k\, A \sin^2\! \theta \,
{p_{\in}^2 \left( 2p_{\in}\cos \theta -k \right)^2 \over \Omega_{\in}\Omega_{\rm f}}
\delta \left( \Omega_{\in} - \Omega_{\rm f} - \omega_{\rm V} \right) \,.
\ea
Since the delta function in the above equation is the same expression as in the tensor case, 
we get the same condition (\ref{cosc}) by replacing $\omega_1$ to $\omega_{\rm V}$.
Assuming $M/p_\in \ll 1$, 
the condition can be rewritten by solving the quadratic inequality as 
\ba \label{kminv}
1 - { \omega_{\rm V} \over k}  \geq {M^2 \over 2p_\in (p_\in-k)}\,.
\ea
The condition for the vector GCR emission is therefore 
given by $1-c_{\rm V} \simgt M^2/p_\in^2$.
For the same reason as in the tensor case, 
we can safely ignore $M^2 / p_\in^2$ in the present paper.
Then, {Eq.}~\eqref{kminv} determines the lower limit of the integration,
\ba
\kminv \equiv {m_{\rm V}\over \sqrt{ 1-c^2_{\rm V} }},
\ea
and thus Eq.~\eqref{A} can be written as 
\ba
A= {\kappa \xi_c^2 \over 1+\kappa \xi_c^2 } { \mu^2  \over k^2} \left[ 1- \left(1-c^2_{\rm V}\right) 
\left( 1 - {\kminv^2 \over k^2} \right) \right]^{-1} \,.
\ea
Then, we find the contribution from $k \simeq \kminv$ in Eq.~(\ref{EV}) is not dominant.
Assuming $k \gg \kminv$, we finally obtain an approximation to  the energy emission rate of the vector GCR, 
\ba
{dE_{\rm V} \over dt} &\simeq& {1\over 4\pi M_g^2 } {\kappa \xi_c^2 \over 1+ \kappa \xi_c^2} \, p_\in^2 \mu^2 (1-c^2_{\rm V})
\int^{p_{\in}}_{\kminv} dk {1 \over k}
\(1-{k \over p_\in}c_{\rm V}-{k^2\over 4p_\in^2}(1-c_{\rm V}^2)\) \(1-{k \over 2p_\in}c_{\rm V}\)^2
\nonumber\\
&\simeq& 
{1\over 4\pi M_g^2}  
{\kappa \xi_c^2 \over 1+ \kappa \xi_c^2} p_\in^2  \mu^2 (1-c^2_{\rm V})\ln \({p_\in \over \kminv}\).
\label{dEdtV}
\ea
Here, we used $m_{\rm V}^2 \simeq \mu^2$ at low energies, and we only kept the leading contribution 
for $p_\in \gg \kminv$ in the last line.

\section{Constraints from high energy cosmic rays}
In this section we derive the condition that the damping due to GCR is not
significant for an ultrahigh energy cosmic ray with initial 
energy $p_\in$ during time $t$, {\it i.e.,} the condition that
$dE_{\rm total}/dt<p_\in /t$ is satisfied,
where $E_{\rm total}=E_{\rm T}+E_{\rm V}+E_{\rm S}$. 
Because of the complexity of the scalar perturbation, 
we only focus on the vector and tensor GCR discussed in Sec. III. 
We assume that 
the origins of high energy cosmic rays are located at a cosmological distance,
${ct\gtrsim} 1~ {\rm Mpc}$, and the initial momentum of the high energy cosmic rays of our concern is $p_\in \sim 10^{11}~ {\rm GeV}$.

Let us first examine the cosmological solution introduced in Sec.\,II. 
The deviation of $\tildec$ from unity is related to 
the effective graviton mass $\mu$ through {Eq.}~(\ref{tildec})
as $1-\tildec \sim H_0^2/\mu^2$. 
Then, the condition for the tensor GCR to occur can be simply given by 
$\mu \simlt 10^{-31} {\rm GeV}$.
In this case we always have $k_D< p_\in$, and
the energy emission rate of the tensor GCR is therefore given by Eq.~(\ref{dEdtT1}). 
Assuming $\kxi \sim {\cal O}(1)$, we have 
\ba
{dE_{\rm T} \over dt} \sim {p_\in^2 H_0^2 \over M_g^2} \ll {p_\in \over t}.
\ea
Since the energy loss of a high energy cosmic ray 
due to the tensor GCR is extremely small,
there is no conflict with observations, {as we have anticipated earlier.}

Let us next consider the vector GCR. 
Assuming $\kxi \sim {\cal O}(1)$ and $\ln(p_\in / \kminv)\sim {\cal O}(1)$, we have 
\ba
{dE_{\rm V} \over dt} \sim {p_\in^2 \mu^2 \over M_g^2} (1-c^2_{\rm V}).
\ea
The constraint on the effective graviton mass is now given by
\ba
\mu \simlt 100 \, (1-c^2_{\rm V})^{{-1/2}}\, {\rm eV},
\ea
which {will allow the whole range of} the graviton mass of our interest. 
Therefore the ghost-free {background} solutions introduced in Sec. II 
{are} consistent with observations of high energy cosmic rays.

\begin{figure}[t]
	\begin{tabular}{cc}
		\begin{minipage}{0.5\textwidth}
			\begin{center}
				\includegraphics[scale=0.6]{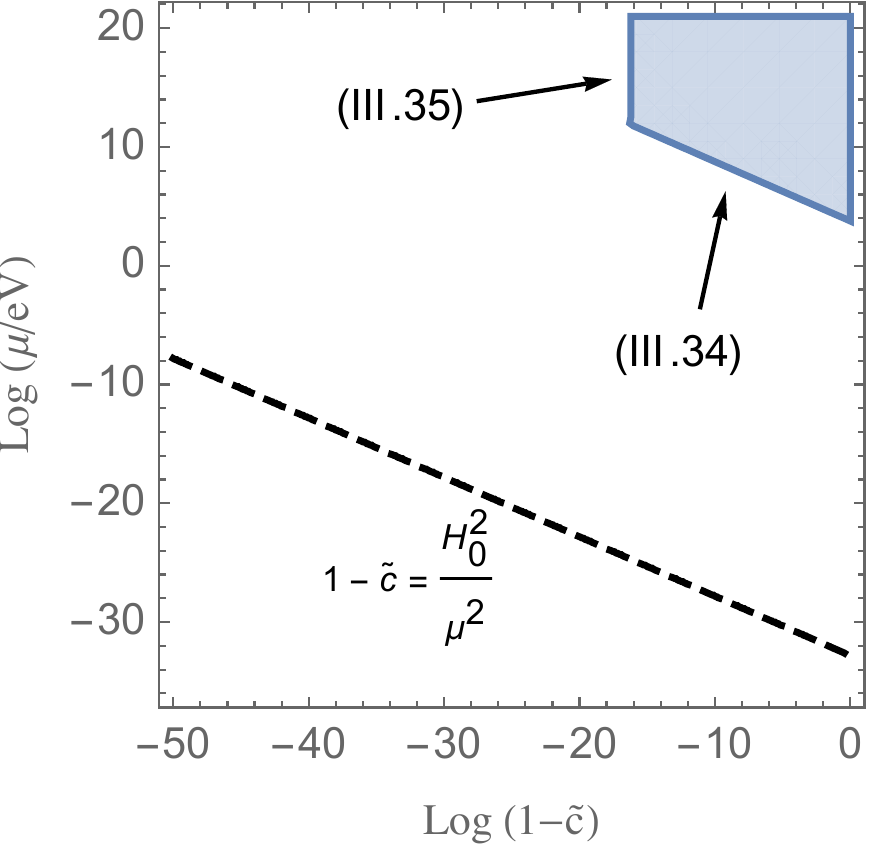}
			\end{center}
		\end{minipage}
		\begin{minipage}{0.5\textwidth}
			\begin{center}
				\includegraphics[scale=0.6]{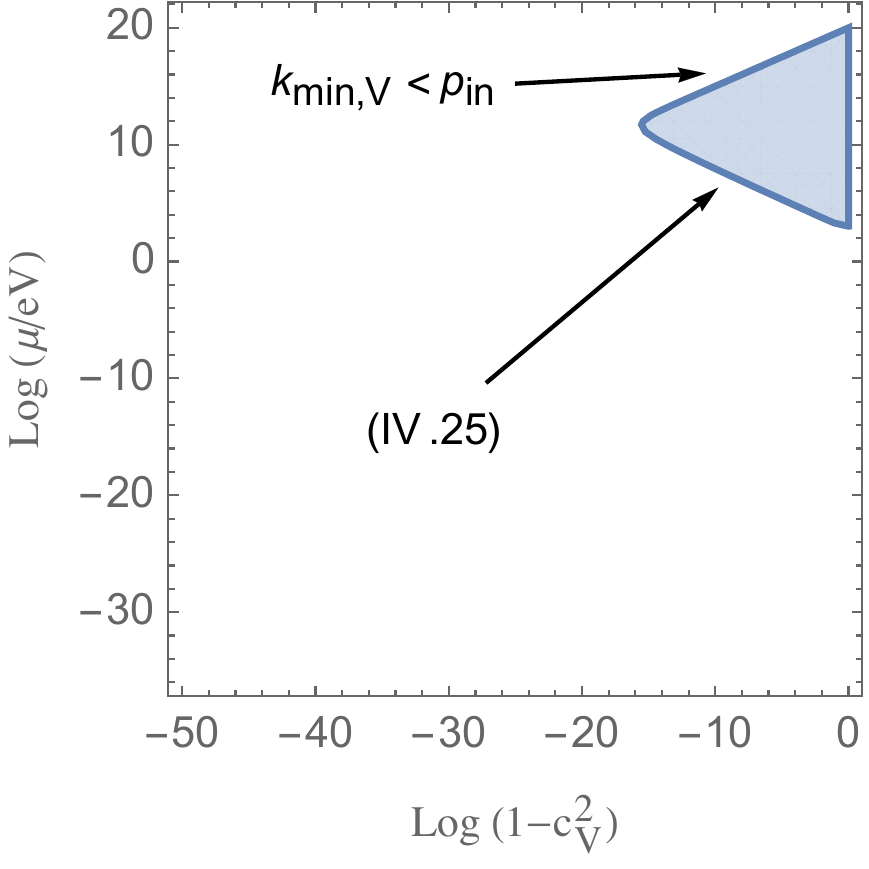}
			\end{center}
		\end{minipage}
	\end{tabular}
	\caption{
		Left : The excluded region in the $\mu$-$(1-\tildec)$ plane obtained by the constraint from the tensor GCR.
		The black dashed line shows the line $(1-\tildec) = H_0^2/\mu^2$, and $\kxi=1$.
		Right :  The excluded region in the $\mu$-$(1-c_{\rm V}^2)$ plane obtained by the constraint from the vector GCR
		for $\kxi=1$.
	}
	\label{fig:constraint}
\end{figure}

Owing to the relation $1-\tildec \sim H_0^2/\mu^2$, 
the tensor GCR emission is suppressed by 
$H_0^2 / M_g^2$. Relaxing this relation, we 
now consider the constraint on $\tildec$ and $\mu$ assuming as if they could be independently determined. 
The shaded region in the left panel of Fig.~\ref{fig:constraint} shows
the excluded region in the $\mu\,$-$\,(1-\tildec)$ plane obtained by the
constraint from the tensor GCR. 
The lower and the left boundaries are, respectively, determined by {the estimates of} 
the emission rate~(\ref{dEdtT1}) and (\ref{dEdtT2}).
One can see that the cosmological solution in Sec.~II, 
which lies at $1-\tildec \sim H_0^2/ \mu^2$ (black dashed line), 
is far from the excluded region. 
Even if we independently treat $1-\tildec$ and $\mu$, 
the tightest constraint on the effective graviton
mass is $\mu \simlt 100 \, {\rm eV}$.
In the right panel of Fig.~\ref{fig:constraint}, we present the excluded region in the $\mu\,$-$\,(1-c_{\rm V}^2)$ plane 
obtained by the constraint from the vector GCR.
The lower and the upper boundaries are, respectively, determined by
the emission rate~(\ref{dEdtV}) and {the condition} $\kminv < p_\in$.
Also {in} the vector case the tightest constraint on the effective graviton mass is $\mu \simlt 100 \, {\rm eV}$ similarly to the tensor case.
Hence, this model is consistent with observations {of ultrahigh} energy cosmic rays.

\section{Conclusion}

In this paper, we studied the consistency of the ghost-free bigravity
model with observations of 
ultrahigh energy cosmic rays.
The GCR can be emitted from a relativistic particle 
when a phase velocity of graviton is slower than the speed of light.
If such a process is possible, 
a high energy cosmic ray reduces its energy 
during its propagation to the Earth, and 
a subluminal phase velocity of graviton 
could be strongly constrained.
In the ghost-free bigravity model that we considered in this paper \cite{Felice:2014aa}, 
the light speed in the hidden metric becomes subluminal or superluminal and then 
the graviton can possess a subluminal phase velocity. 
We confirmed that a relativistic particle emits
the GCR in this model and derived the conditions for such a process to occur. 
The energy emission rate of the GCR of the tensor mode and the vector
mode was estimated, and it turned out to be suppressed
as far as the effective graviton mass is sufficiently small to satisfy
$\mu \simlt 100 \, {\rm eV}$, which will cover most of the parameter
region that is interesting when we consider gravity modification 
relevant at a late epoch.  

Although we did not derive the emission rate of the scalar GCR 
due to the complexity of the dispersion relations, 
we think it natural to assume the emission rate of the scalar GCR 
is also suppressed for the following reason. 
In Ref.~\cite{2009PThPh.121..427I}
the coupling between the scalar mode 
of a simple FP massive graviton and a real conformal scalar field 
in de Sitter background was computed, and the coupling squared, 
which is proportional to the transition amplitude,  
was reported to be suppressed by the factor 
$\mu^2(\mu^2-2H^2)/\tilde M^2_g k^2$ (We use $\tilde M^2_g$ instead of $M^2_g$ since the only option here is the effective gravitational constant in the context of the FP massive graviton.). 
Although this computation was done not in the context of 
bigravity without taking into account 
the coupling between 
$g$- and $f$- matters and graviton, 
the factor mentioned above is, in a naive sense, 
the quantity to be compared with 
the vector mode counterpart 
\begin{equation}
\frac{\kappa\xi_c^2}{1+\kappa\xi_c^2}
\frac{k^2 A}{M_-^2}
\approx 
\frac{\kappa\xi_c^2}{1+\kappa\xi_c^2}
\frac{\mu^2}{c_{\rm V}^2 M_g^2}\,,
\end{equation}
in the present setup. Neglecting the factor related to 
$\kappa\xi_c^2$ and the deviation of $c_{\rm V}^2$ from unity, 
we find that the coupling between the scalar mode of massive graviton
and the incident high energy particle is as suppressed  
as in the case of the vector mode. 
On the other hand, 
the propagation speed of the scalar mode of graviton in bigravity 
has been calculated in Ref.~\cite{Felice:2013aa}, and the 
obtained expression is 
similar to the vector case at low energies (Eq.~(88) in \cite{Felice:2013aa}). 

When we consider non-conformal field, we need to keep 
the trace-part of the metric perturbation, which was 
neglected in the computation in Ref.~\cite{2009PThPh.121..427I}.
This neglected contribution gives a coupling to the trace-part 
of the energy momentum tensor, which is absent as long as we 
consider conformally invariant matter fields. 
As is expected from the presence of the vDVZ discontinuity, 
the trace-part of the energy momentum tensor 
will couple to the scalar mode of massive graviton without 
any suppression even in the massless limit. 
However, such a non-conformal component of the matter 
energy momentum tensor will be suppressed by 
the degree of violation of the conformal invariance, {\it i.e.}, 
by the ratio of mass to momentum squared, $M^2/p_{\rm in}^2$, in 
the case of a Dirac fermion, instead 
of the suppression by $\mu^2/p_{\rm in}^2$. 
As a result, the transition amplitude should have a 
suppression factor proportional to $(M^2/p_{\rm in}^2)^2$. 
Then, based on the dimensional argument, the GCR emission rate would 
be, at most, given by $dE/dt\approx M^4/M_g^2$. 
When ultrahigh energy cosmic ray protons are concerned, 
the fraction of the energy that is lost by the GCR after 
traveling a cosmological distance is much less than unity. 
Therefore, we expect that the scalar GCR will be harmless 
and conclude that the ghost-free bigravity with a sufficiently small 
mass is consistent with the observations of high energy cosmic rays, 
although confirmation by an explicit computation for scalar mode
is needed to obtain a conclusive answer.

 \acknowledgments
RK is supported by the Grant-in-Aid for Japan Society for the Promotion of Science (JSPS) 
Grant-in-Aid for Scientific Research Nos. 25287054. 
TT is supported by the Grant-in-Aid for Scientific
Research (Nos. 24103006, 24103001, 26287044 and 15H02087). 
YY is supported by the Grant-in-Aid for JSPS Fellows No. 15J02795. 
The research by K.Y. is supported by a Grant-in-Aid for Scientific Research of Japan Ministry of Education, 
Culture, Sports, Science and Technology (No.15H05895).

\bibliography{references}

\end{document}